\documentclass{article}
\usepackage{spconf,amsmath,graphicx}

\usepackage{multirow,booktabs,paralist,bbding,pifont,graphicx}

\title{SA-SOT: Speaker-Aware Serialized Output Training for Multi-Talker ASR}
\name{Zhiyun Fan, Linhao Dong, Jun Zhang, Lu Lu, Zejun Ma}
\address{
  Bytedance Research \\
  \{fanzhiyun.20203,donglinhao\}@bytedance.com}
\begin{document}
\ninept
\maketitle
\begin{abstract}
Multi-talker automatic speech recognition plays a crucial role in scenarios involving multi-party interactions, such as meetings and conversations. Due to its inherent complexity, this task has been receiving increasing attention. Notably, the serialized output training (SOT) stands out among various approaches because of its simplistic architecture and exceptional performance. However, the frequent speaker changes in token-level SOT (t-SOT) present challenges for the autoregressive decoder in effectively utilizing context to predict output sequences. To address this issue, we introduce a masked t-SOT label, which serves as the cornerstone of an auxiliary training loss. Additionally, we utilize a speaker similarity matrix to refine the self-attention mechanism of the decoder. This strategic adjustment enhances contextual relationships within the same speaker's tokens while minimizing interactions between different speakers' tokens. We denote our method as speaker-aware SOT (SA-SOT). Experiments on the Librispeech datasets demonstrate that our SA-SOT obtains a relative cpWER reduction ranging from 12.75\% to 22.03\% on the multi-talker test sets. Furthermore, with more extensive training, our method achieves an impressive cpWER of 3.41\%, establishing a new state-of-the-art result on the LibrispeechMix dataset.

\end{abstract}
\section{Introduction}
\label{sec:intro}

Speaker overlap frequently occurs in scenarios like meetings or conversations \cite{fiscus2007rich,mccowan2005ami,watanabe2020chime,raj2021integration,yu2022m2met}. Traditional single-talker speech recognition systems face challenges in accurately deciphering the content from each speaker \cite{kanda2020joint}. This highlights the importance of the multi-talker speech recognition task, which aims to improve the precision of transcribing overlapped speech instances.

Numerous efforts have been dedicated to addressing the challenge of multi-talker automatic speech recognition (ASR). A promising strategy entails the utilization of a cascaded framework, which integrates a speech separation front-end with an ASR back-end  \cite{chang2020end,chang2019mimo,scheibler2023end}. Notably, Change et al. \cite{chang2019mimo} employ a multi-source neural beamformer to handle multi-channel inputs, coupled with a multi-output ASR configuration to yield individualized output for each speaker. The entire model is optimized using permutation invariant training (PIT) \cite{qian2018single,seki2018purely} ASR loss. Because the cascade structure is complex and not conducive to stream modeling \cite{moriya2022streaming}, some researchers have explored eliminating the front-end separation module. Instead, they directly input overlapped speech into the ASR model, incorporating speaker differentiation into the encoder's structural design. This modified approach retains the use of multiple outputs in conjunction with the PIT loss at the decoder  \cite{chang2019end,tripathi2020end,sklyar2021streaming,guo2021multi}. Although these tasks are simpler in inputs, the use of multi-output and PIT loss still have several limitations. Firstly, the model's capacity to handle speakers is restricted by the number of output branches \cite{kanda2020serialized}. Secondly, calculating the PIT loss necessitates consideration of all possible combinations of hypotheses and labels, leading to a significant increase in computational complexity as the number of speakers rises \cite{dovrat2021many}. To overcome these constraints, Kanda et al. propose the serialized output training (SOT)-based multi-talker ASR model \cite{kanda2020serialized}. The SOT approach introduces a special symbol, denoted as $<$cc$>$, to indicate the change of the speaker channel and consolidates multiple label sequences from multiple speakers into a unified label sequence at either the sentence level \cite{kanda2020serialized} or token level \cite{kanda2022streaming}, aligning with the chronological order of sentences or tokens. This adaptation enables the training of multi-talker ASR tasks utilizing the same model architecture and training loss as single-talker ASR. Compared with alternative methods, SOT not only has a simple structure but also demonstrates remarkable efficacy. 

\begin{figure}[t]
  \centering
  \includegraphics[width=0.9\linewidth]{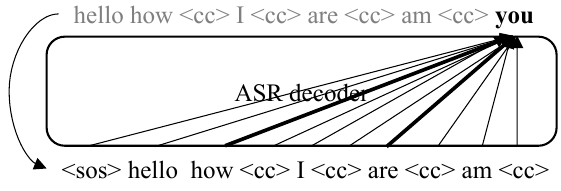}
  \caption{An example of the decoding process of t-SOT-based ASR model. For simplicity, the encoder is omitted. The full t-SOT label is 'hello how $<$cc$>$ I $<$cc$>$ are $<$cc$>$ am $<$cc$>$ you $<$cc$>$ fine'.}
  \vspace{-0.4cm}
  \label{fig:example}
\end{figure}

Although the SOT-based method simplifies the model structure and training with the introduction of the special symbol $<$cc$>$, it also brings new challenges. Firstly, during the inference stage, the model's performance becomes significantly dependent on the accurate prediction of $<$cc$>$, which is closely tied to speaker changes. However, ASR itself is inherently a speaker-independent task \cite{liang2023ba,adi2019reverse}. To address this concern, Liang et al. \cite{liang2023ba} propose the boundary-aware SOT (BA-SOT), incorporating speaker change detection (SCD) block and a boundary constraint loss to assist the model in predicting $<$cc$>$, and the performance of the SOT approach has been significantly improved. Secondly, the consolidation of multiple label sequences into a single sequence heightens the complexity of learning for the autoregressive decoder, which predicts subsequent tokens based on contextual history. In this paper, we refer to this challenge as a semantic confusion problem. This problem becomes particularly serious in the token-level serialized output training (t-SOT) methodology \cite{kanda2022streaming}. An illustration of this challenge is provided in Fig. \ref{fig:example}. When the ASR decoder predicts the word 'you', it must discern that 'hello how are' constitutes valid historical context, while 'I am' represents interjections from a different speaker.

To address this challenge, we propose speaker-aware serialized output training (SA-SOT). Our method involves multi-task training encompassing both ASR and speaker classification. In contrast to \cite{kanda2020joint}, our speaker classification doesn't rely on predefined profiles. Moreover, distinct from the approach in \cite{kanda2022streaming2}, which solely employs speaker representation post-processing for labeling speakers, we integrate the extracted speaker embeddings into the ASR decoder to enhance the SOT. What's more, the speaker embeddings are used to generate a similarity matrix which is used to modify the self-attention within the ASR decoder. This modification strengthens the contextual relationships among the same speaker's tokens while reducing the influence of contextual relationships between different speakers. Furthermore, we employ a masked t-SOT label to guide the ASR decoder in identifying which tokens belong to the same speaker. Our experiments, conducted on the Librispeech data set, demonstrate the effectiveness of SA-SOT in improving the performance of SOT-based models. We obtain a remarkable 12.75\% to 22.03\% relative reduction in the concatenated minimum-permutation word error rate (cpWER) on the multi-talker test sets, while maintaining the performance on the single-talker test set. Moreover, with more extensive training, we obtain 3.41\% cpWER, establishing a new state-of-the-art (SOTA) result on the LibrispeechMix test set. In summation, we conclude our contributions as follows:

\begin{compactitem}
    \item  We have validated the efficacy of the t-SOT method on the continuous integrate-and-fire (CIF)-based ASR model \cite{dong2020cif}, achieving results that are on par with existing research.
    \item Our integration of token-level speaker embeddings into the ASR decoder has yielded significant performance improvements.
    \item We have introduced the innovative concept of utilizing a speaker similarity matrix to adapt the self-attention mechanism within the ASR decoder.
    \item We propose the use of masked t-SOT labels to guide the model in accurately attributing tokens to specific speakers.
\end{compactitem}

\section{Proposed Method}
\label{sec:format}
In this paper, we present our method within a framework of joint modeling of automatic speech recognition and speaker classification. Our primary focus in this study lies on the ASR branch, particularly its performance in the multi-talker speech scenarios. Concurrently, the speaker branch operates as an auxiliary task, producing valuable speaker embeddings. The core of our methodology revolves around leveraging speaker embeddings to enhance the efficacy of the t-SOT-based ASR, especially when dealing with the complexities of multi-talker speech.
\begin{figure}[t]
  \centering
  \includegraphics[width=\linewidth]{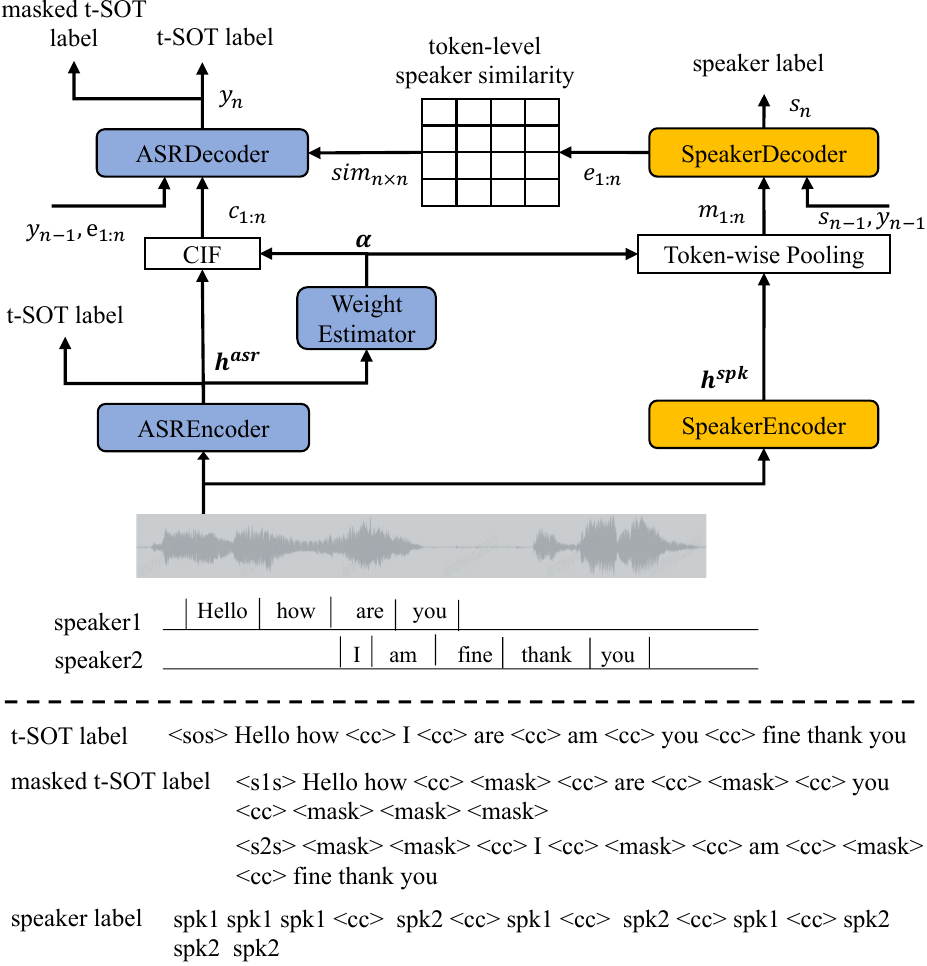}
  \caption{The upper of the figure is the overview of our SA-SOT-based method. ASR branch is colored by blue, the speaker  branch is colored by yellow. The lower part is an example showing the labels used in the training.}
  \vspace{-0.4cm}
  \label{fig:framework}
\end{figure}
\vspace{-0.1cm}
\subsection{Framework}
\vspace{-0.1cm}

The details of the model framework are illustrated in Fig. \ref{fig:framework}. The ASR branch employs a CIF-based encoder-decoder structure. The ASR encoder transforms the input feature $\boldsymbol{x}$ into the encoded outputs $\boldsymbol{h^{asr}}$. The weight estimator evaluates the acoustic information for each frame, generating frame-level information weight $\boldsymbol{\alpha}$. Subsequently, the CIF module utilizes $\boldsymbol{\alpha}$ to convert the frame-level encoded outputs $\boldsymbol{h}$ into token-level acoustic embeddings $\boldsymbol{c}$. It should be known that the acoustic boundaries are predicted during the calculation of the CIF module. For more details of the CIF module, we refer readers to \cite{dong2020cif}. The ASR decoder uses $\boldsymbol{c}$ to predict the ASR label sequence in an autoregressive manner. The computation process for the ASR branch can be formulated as follows:
\begin{equation}
\setlength{\abovedisplayskip}{3pt} 
\setlength{\belowdisplayskip}{3pt}
h_{[1:T']}^{asr}=\text{AsrEncoder}(x_{[1:T]})
\end{equation}
\begin{equation}
\setlength{\abovedisplayskip}{3pt} 
\setlength{\belowdisplayskip}{3pt}
\alpha_{[1:T']}=\text{WeightEstimator}(h_{[1:T']}^{asr})
\end{equation}
\begin{equation}
\setlength{\abovedisplayskip}{3pt} 
\setlength{\belowdisplayskip}{3pt}
c_{[1:N]}=\text{CIF}(h_{[1:T']}^{asr},\alpha_{[1:T']})
\end{equation}
\begin{equation}
\setlength{\abovedisplayskip}{3pt} 
\setlength{\belowdisplayskip}{3pt}
y_{n}=\text{AsrDecoder}(c_{[1:n]},\text{Embed}(y_{[1:n-1]}))
\label{eq: asrdec raw}
\end{equation}
where $N$ represents the number of predicted tokens, $T$ is the length of input feature sequence, $T'$ is the length of encoded output after downsampling.

The speaker branch comprises an encoder, a decoder, and a token-wise average pooling layer. The encoder converts the input feature to the frame-level speaker embedding $\boldsymbol{h^{spk}}$. The token-wise average pooling layer utilizes the acoustic boundary from the ASR branch to average these frames between two continuous acoustic boundaries and generate the token-level speaker embedding $\boldsymbol{m}$. The speaker decoder uses $\boldsymbol{m}$ to autoregressively predict the speaker label sequence. In our initial experiments, we encountered challenges when attempting to directly predict speakers for overlapping speech instances. As a solution, we incorporate ASR outputs $y_{n-1}$ into the speaker decoder. The computation process for the speaker branch can be summarized as follows:
\begin{equation}
\setlength{\abovedisplayskip}{3pt} 
\setlength{\belowdisplayskip}{3pt}
h_{[1:T']}^{spk}=\text{SpeakerEncoder}(x_{[1:T]})
\end{equation}
\begin{equation}
\setlength{\abovedisplayskip}{3pt} 
\setlength{\belowdisplayskip}{3pt}
m_{[1:N]}=\text{T-AveragePooling}(h_{[1:T']}^{spk})
\end{equation}
\begin{equation}
\setlength{\abovedisplayskip}{3pt} 
\setlength{\belowdisplayskip}{3pt}
s_{n}=\text{SpeakerDecoder}(m_{[1:n]},s_{[0:n-1]},y_{[0:n-1]})
\end{equation}

\vspace{-0.1cm}
\subsection{Speaker embedding fusion}
\vspace{-0.1cm}

As ASR inherently operates as a speaker-independent task \cite{adi2019reverse}, it's challenging for an ASR model to predict all tokens for each speaker in overlapped speech. Consequently, we consider utilizing speaker information as an auxiliary. The output of the speaker decoder's hidden layer $\boldsymbol{e}$ is used as speaker embedding, which is then incorporated into the ASR decoder.
\begin{equation}
\setlength{\abovedisplayskip}{3pt} 
\setlength{\belowdisplayskip}{3pt}
y_{n}=\text{AsrDecoder}(c_{[1:n]},\text{Proj}(\text{Embed}(y_{[0:n-1]}), e_{[1:n]}))
\label{eq: asrdec mod}
\end{equation}
where $\text{Embed}(\cdot)$ is an embedding function. $\text{Proj}(\cdot)$ presents a projection layer that is used to ensure that the dimensions of the ASR decoder input are appropriate. We refer to this operation as speaker embedding fusion (SEF), and the Eq. \ref{eq: asrdec raw} will be turned to Eq. \ref{eq: asrdec mod}.

\vspace{-0.1cm}
\subsection{Speaker-aware training}
\vspace{-0.1cm}
The t-SOT label is shown on the lower part of Fig. \ref{fig:framework}. It is noticeable that the special symbol $<$cc$>$ appears frequently in the overlapped speech to signify the speaker changes. For the autoregressive ASR decoder, modeling context in such a scenario becomes challenging. To address this issue, we propose a masked t-SOT label, as shown in Fig. \ref{fig:framework}. For specific speakers, we employ a special token $<$mask$>$ to mask the tokens from other speakers. And different from the t-SOT label, we utilize special start symbols $<$s1s$>$ or $<$s2s$>$ to indicate that this is an auxiliary training loss. Our inference begins with the $<$sos$>$ token. We refer to the cross-entropy (CE) loss associated with masked t-SOT labels as speaker-aware training (SAT) loss. 

\vspace{-0.1cm}
\subsection{Speaker-aware attention}
\vspace{-0.1cm}
In the previous section, we utilize the speaker-aware training loss to emphasize that specific tokens belong to the same speaker. To further strengthen the contextual relationships within the same speaker and reduce contextual relationships between different speakers, we modify the self-attention mechanism of the ASR decoder.

As illustrated in Fig. \ref{fig:framework}, we derive an $N$ by $N$ similarity matrix by computing the cosine similarity between pairs of token-level speaker embeddings $e_{[1:N]}$.
\begin{equation}
\setlength{\abovedisplayskip}{3pt} 
\setlength{\belowdisplayskip}{3pt}
sim_{i,j}=\text{cos}(e_i,e_j)
\end{equation}

The original self-attention in the ASR decoder is expressed as:
\begin{equation}
\setlength{\abovedisplayskip}{3pt} 
\setlength{\belowdisplayskip}{3pt}
\text{Attention}(Q,K,V)=\text{softmax}(\frac{QK^T}{\sqrt{d_k}})V
\label{eq:raw att}
\end{equation}

We use the similarity matrix to modify self-attention, which we refer to as speaker-aware attention (SAA). 
\begin{equation}
\setlength{\abovedisplayskip}{3pt} 
\setlength{\belowdisplayskip}{3pt}
\text{weight}=\text{softmax}(\frac{QK^T}{\sqrt{d_k}})\odot\frac{1+sim}{2}
\label{eq: we}
\end{equation}

\begin{equation}
\setlength{\abovedisplayskip}{3pt} 
\setlength{\belowdisplayskip}{3pt}
\text{O}=(\text{weight}/\sum_{j}{\text{weight}_{[:,j]}})V
\label{eq:o}
\end{equation}
where the $\odot$ denotes the Hadamard product. The similarity value is scaled to interval $(0,1)$ and is used to reweight the original attention weight in Eq \ref{eq: we}. In Eq. \ref{eq:raw att}, the weight is renormalized and applied to weight $V$. By implementing the proposed SAA, Eq. \ref{eq:raw att} is turned to Eq. \ref{eq: we} and Eq. \ref{eq:o} within the self-attention of the ASR decoder.

\vspace{-0.1cm}
\subsection{Training loss}
\vspace{-0.1cm}
We employ a joint training strategy to optimize all parameters, where the overall objective function is a weighted sum of all the losses. In the speaker branch, AMsoftmax loss \cite{liu2019large} is employed to capture more discriminative speaker embeddings. In the CIF-based ASR branch,  in addition to the CTC loss, CE loss, and quantity loss used in the original CIF-based model, the auxiliary SAT loss is also incorporated.
\begin{equation}
\setlength{\abovedisplayskip}{3pt} 
\setlength{\belowdisplayskip}{3pt}
\mathcal{L}=\mathcal{L}_{ce}+\lambda_1 \mathcal{L}_{ctc}+\lambda_2\mathcal{L}_{qua}+\lambda_3\mathcal{L}_{ams}+\lambda_4\mathcal{L}_{sat}
\end{equation}
where the  $\mathcal{L}_{ce}$, $\mathcal{L}_{ctc}$, and $\mathcal{L}_{qua}$ are the same as in the vanilla CIF-based model \cite{dong2020cif}. The $\mathcal{L}_{ams}$ represents AMsoftmax loss \cite{liu2019large}. The $\mathcal{L}_{sat}$ represents the SAT loss which employs the same formulas as $\mathcal{L}_{ce}$ but utilizes the masked t-SOT label. $\lambda_1$ to $\lambda_4$ are tunable hyper-parameters.

\section{Experiments}
\label{sec:exp}
\vspace{-0.1cm}
\subsection{Datasets and metric}
\vspace{-0.1cm}
All our experiments are conducted on the Librispeech corpus \cite{kanda2020joint} and LibrispeechMix \cite{kanda2020joint}. We use the 960-hour training data of Librispeech to simulate multi-talker training data, and the test-clean and test-other sets to simulate two multi-talker test sets (test-clean-2mix/test-other-2mix) following the same process. Take the training set as an example, each sample in the 960-hour training set overlaps with another sample with a probability $p$, and remains unchanged with a probability of $1-p$. The other samples used to simulate overlap are also randomly chosen from the 960-hour training set. The waveform of the two samples are then added together with a random shift of start time. The labels of two speakers are merged together in a chronological order based on the word emission time. And a special symbol $<$cc$>$ is used to indicate the speaker changes. The word emission time is generated by the Montreal Forced Aligner \cite{mcauliffe2017montreal}.

The LibrispeechMix is a test set containing two speakers in each speech, simulated through a process similar to our training dataset. To observe the impact of our method on samples without speaker overlap, we also present the experimental results for the test-clean of Librispeech. As for metric, we adapt the concatenated minimum-permutation word error rate (cpWER) \cite{watanabe2020chime}.

\vspace{-0.1cm}
\subsection{Experimental Setup}
\vspace{-0.1cm}
For all experiments, we utilize 80-dimensional log-Mel filter-bank features, which are computed with a 25 ms window and shifted every 10 ms. Two convolutional layers with 128 filters and 1/4 temporal downsampling are used as the front end of the ASR encoder. The ASR encoder comprises 18 conformer \cite{gulati2020conformer} layers with 8 attention heads, 512 attention dimensions, and 2048 feed-forward network (FFN) dimensions. The weight estimator consists of a 1-dimensional convolutional layer with a kernel size of 3 and 512 filters, followed by a fully-connected (FC) layer with a single output unit using sigmoid activation. The ASR decoder consists of 2 transformer \cite{vaswani2017attention} layers with 8 attention heads, 512 attention dimensions, and 2048 FFN dimensions. The speaker encoder has the same architecture as the 18-layer ResNet in \cite{he2016deep} except for not the final average pooling layer and halving the channels of all the convolutional layers. Upon the ResNet, two conformer layers are used to capture global features, which are of the same settings as ASR encoder layers. Both the speaker encoder and the ASR encoder have a 1/4 temporal downsampling. The speaker decoder stacks two FC layers. The first layer has 256 units with ReLU activation, and the second layer has 2341 (total speaker number in the training set) output units with softmax nonlinearity. The threshold $\beta$ used in the CIF module is set to 1. The modeling units comprise 3730 byte pair encoding (BPE) subwords. The hyper-parameters $\lambda_1$, $\lambda_2$, $\lambda_3$, and $\lambda_4$ are set to 0.5, 1.0, 0.1, 1.0, respectively.

\begin{table*}[t]
  \caption{The performance (cpWER \%) of our proposed method and all baseline on the test sets. Results from previous papers and our implementation of the CIF t-SOT model are shown for comparison. The values in parentheses represent the relative cpWER gains.}
  \label{tab:base}
  \centering
  \tabcolsep=0.3cm
  \begin{tabular}{llccccc} 
    \toprule
\textbf{ID} &\textbf{Model} & Model size &  test-clean & LibrispeechMix & test-clean-2mix & test-other-2mix   \\
%\cmidrule(lr){2-3}\cmidrule(lr){4-5}
      \midrule 
    B1 &  SOT LSTM SA-ASR  \cite{kanda2020joint} & 146M &  4.2 & 8.7 & - &-  \\
    B2 &  SOT Confomer SA-ASR \cite{kanda2021end} & 142M &  3.3 & 4.3 & - &-  \\
     B3 & t-SOT TT \cite{kanda2022streaming2}  & 139M & 3.3 & 4.4 & - & - \\
      \midrule
    E1 &     CIF t-SOT      &   136M & 2.94 & 5.10 &  5.22 & 6.57 \\
    E2 &  \ \   + SEF       &   136M & 2.89 (1.70\%) & 4.79 (6.08\%) & 4.55 (12.84\%)& 6.14 (6.54\%) \\
    E3 & \ \ \ \    + SAT   &   136M & 2.88 (2.05\%) & 4.55 (10.78\%) & 4.15 (20.50\%)& 5.52 (15.98\%)\\
    E4 &   \ \ \ \ \ \   + SAA & 136M & 2.90 (1.36\%) & 4.45 (12.75\%) & 4.07 (22.03\%) & 5.52 (15.98\%)\\
     E5 &  \ \ \ \ \ \ \ \  + Extensive Training & 136M & 2.52 (14.29\%) & 3.41 (38.43\%) & 1.97 (62.26\%) & 3.27 (47.18\%) \\

   \bottomrule
  \end{tabular}
\end{table*}

\begin{table}[t]
  \caption{Impact of the encoder downsampling on cpWER(\%) for our proposed method.}
  \label{tab:downsampling}
  \centering
  \tabcolsep=0.4cm
  \begin{tabular}{lccc} 
    \toprule
Downsampling & 1/2 & 1/4 & 1/8  \\
      \midrule 
test-clean &  3.19 &2.90 &  2.93\\
LibrispeechMix & 4.62  &4.45 & 6.46\\
test-clean-2mix & 4.06 & 4.07& 5.85\\
test-other-2mix & 5.49 & 5.52& 7.22\\
%\cmidrule(lr){2-3}\cmidrule(lr){4-5}
   \bottomrule
  \end{tabular}
\end{table}

In the training stage, we utilize the Adam optimizer \cite{kingma2014adam}. The ratio of overlap samples $p$ is set to 0.5. The learning rate warms up for the initial 1000 iterations to reach a peak of 1e-3 and holds on for the subsequent 40k iterations, and then linearly decays to 1e-4 for the final 60k iterations. The last 10 checkpoints are averaged to obtain the final model for evaluation. For decoding, we employ beam-search \cite{graves2012sequence} with a beam size of 10.
\vspace{-0.1cm}
\subsection{Main results}
\vspace{-0.1cm}
Table \ref{tab:base} provides a comprehensive comparison between our model and several strong baseline models on the test-clean and the three multi-talker test sets. Baselines B1 to B3 are SOT-based transducer or AED method. E1 is the CIF-based model trained with t-SOT, and we adjust the parameters of our models to closely match the baseline models. Impressively, CIF t-SOT obtains comparable results with the baseline models. From E1 to E4, we gradually augment our approach with speaker embedding fusion (SEF), speaker-aware training (SAT), and speaker-aware attention (SAA). The cpWER results on LibrispeechMix demonstrate that incorporating SEF, SAT, and SAM leads to relative cpWER gains of 6.1\% 5.1\% and 2.1\%, respectively. Finally, comparing E4 with E1, our methods obtain a 12.75\% relative WER gain, and 22.03\% and 15.98\% on test-clean-2mix and test-other-2mix, respectively. It is worth noting that our method can improve the performance of the model on the multi-talker test set while maintaining the recognition performance on the single-talker test-clean set. 

Since some data in our training process is simulated on the fly, we try to increase the training steps for extensive training. We find that until the training steps are increased to 500 thousand, the model doesn't overfit on the development set. Ultimately, we achieved obvious improvements on both single-talker and multi-talker test sets through extensive training. Especially, on the LibrispeechMix test set, we obtain 3.41\% cpWER, which is a new state-of-the-art result on the test set. The improvements for the single-talk test set could be attributed to the data augmentation effect resulting from the use of overlapped speech during training, which was also observed in \cite{kanda2020joint}.
\vspace{-0.1cm}
\subsection{Encoder downsampling}
\vspace{-0.1cm}
Table \ref{tab:downsampling} shows the cpWER of our SA-SOT method with different encoder downsampling. For multi-talker test sets, a high frame rate can provide more details, enabling the model to handle overlapped speech more effectively. This is why 1/2 or 1/4 temporal downsampling, as compared to 1/8 temporal downsampling, brings noticeable benefits to the three multi-talker test sets. On the other hand, we notice that for the single-talker test set the 1/2 temporal downsampling is worse than the other two settings. We attribute this to the fact that excessively long sequences bring additional difficulty to the learning of the model. 
\vspace{-0.1cm}
\subsection{Example analysis}
\begin{figure}[t]
  \centering
  \includegraphics[width=\linewidth]{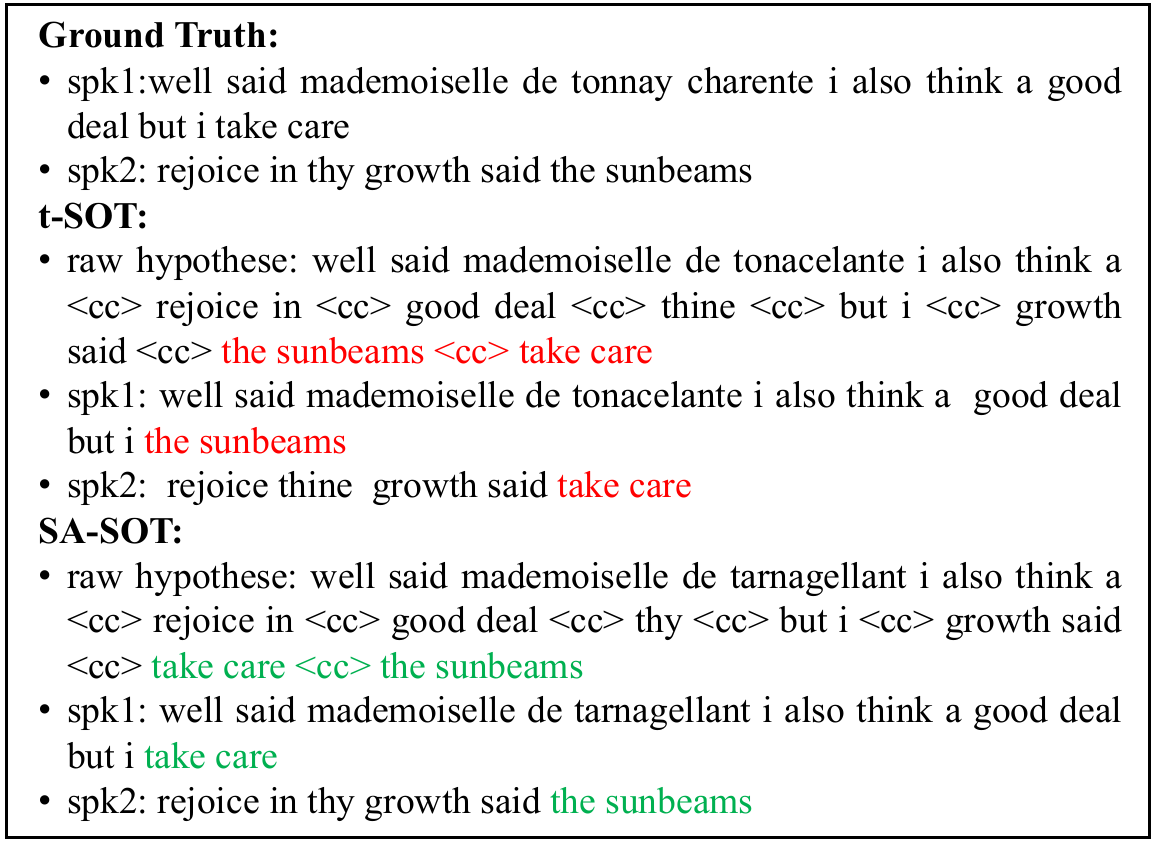}
  \caption{An example of model outputs and reference from the LibrispeechMix. The results for 'spk1' and 'spk2' are obtained by separating the raw hypotheses with the special symbol '$<$cc$>$'.}
  \vspace{-0.4cm}
  \label{fig:demo}
\end{figure}

\vspace{-0.1cm}
An actual sample is illustrated in Figure \ref{fig:demo}. For both t-SOT and our SA-SOT method, the final multi-talk results are separated from the raw hypotheses using the special symbol $<$cc$>$. The t-SOT baseline method misassigns "the sunbeams" and "take care" to the wrong speakers, resulting in four additional errors when calculating the cpWER. In contrast, our SA-SOT method addresses this issue by enabling more accurate speaker assignments as we anticipated.

\section{Conclusion}
In this paper, we propose the speaker-aware serialized output training (SA-SOT) method to address the semantic confusion problem existing in the t-SOT method. In our approach, the token-level speaker embedding is incorporated into the ASR decoder to assist in processing overlapped speech. Additionally, the speaker similarity matrix is utilized to modify the attention weight in the self-attention layer of the ASR decoder to strengthen the contextual relationships of the same speaker and reduce contextual relationships between different speakers. Furthermore, we designed the masked t-SOT label to prompt the model which tokens belong to the same speaker. Trained on the multi-talker data simulated with the Librispeech data set, our method has demonstrated substantial improvements across both multi-talker and single-talker test sets.

%\vfill
%\pagebreak

%\vfill\pagebreak

\bibliographystyle{IEEEbib}
\bibliography{strings,refs}

\end{document}